%% file: masterfile.tex
\documentclass[10pt,twoside,BCOR7mm,DIV15,headinclude,footexclude,
               cleardoubleempty,idxtotoc]
{scrartcl}

\usepackage{natbib}
\usepackage[font=small,labelfont=bf]{caption}
\usepackage[english]{babel}
\usepackage{graphicx}
\usepackage{hyperref}
\usepackage{scrpage2}
\usepackage{ifthen}
\usepackage{booktabs}
\usepackage{amsmath}
\usepackage{amssymb}
\usepackage{multicol}
\usepackage{float}
\usepackage{hyperref}
\usepackage{epstopdf}

\hypersetup{breaklinks=true
,colorlinks=true,linkcolor=black,urlcolor=blue
,citecolor=black}

\addto\captionsenglish{%
}

\makeatletter
\renewcommand{\@biblabel}[1]{}
\renewcommand{\@cite}[2]{%
{#1\ifthenelse{\boolean{@tempswa}}{,#2}{}}}
\makeatother
\ofoot{\thepage}
\ifoot{}

\setcapindent{0em}
\setheadsepline{1pt}

\setkomafont{pagehead}{\normalfont\sffamily}
\setkomafont{pagenumber}{\normalfont\rmfamily}

\makeatletter
\newcommand{\listofcontributions}{\@starttoc{con}}

\newcommand{\l@contribution} {\@dottedtocline{1}{1.5em}{2.3em}}
\makeatother

\newenvironment{contribution}{
\setcounter{section}{0}
\setcounter{figure}{0}
\setcounter{table}{0}
}{
\newpage
\lehead{}
\rohead{}
}

\begin{document}

\setlength{\baselineskip}{2.5ex}

\begin{contribution}
\include{myarticle}

\end{contribution}


\end{document}

%% file: myarticle.tex

\lehead{C.\ M.\ P.\ Russell et al.}

\rohead{Modeling X-rays from colliding WR winds}

\begin{center}
{\LARGE \bf Hydrodynamic and radiative transfer modeling of X-ray emission from colliding WR winds: \hspace{10cm} WR~140 \& the Galactic center}\\
\medskip

{\it\bf C.\ M.\ P.\ Russell$^1$, M.\ F.\ Corcoran$^{1,2}$, J.\ Cuadra$^3$, S.\ P.\ Owocki$^4$, Q.\ D.\ Wang$^5$,\\ K.\ Hamaguchi$^{1,6}$, Y.\ Sugawara$^7$, A.\ M.\ T.\ Pollock$^8$ \& T.\ R.\ Kallman$^1$}\\

{\it $^1$X-ray Astrophysics Laboratory, NASA/GSFC, USA}\\
{\it $^2$University Space Research Association, USA}\\
{\it $^3$Pontificia Universidad Cat$\acute{o}$lica de Chile, Chile}\\
{\it $^4$University of Delaware, USA}\\
{\it $^5$University of Massachusetts Amherst, USA}\\
{\it $^6$University of Maryland, Baltimore County, USA}\\
{\it $^7$Chuo University, Japan}\\
{\it $^8$University of Sheffield, England}\\

\begin{abstract}
  Colliding Wolf-Rayet (WR) winds produce thermal X-ray emission widely observed by X-ray telescopes.  In wide WR+O binaries, such as WR~140, the X-ray flux is tied to the orbital phase, and is a direct probe of the winds' properties. In the Galactic center, $\sim$30 WRs 
  orbit the super massive black hole (SMBH) within $\sim$10", leading to a smorgasbord of wind-wind collisions. 
  To model the X-ray emission of WR~140 and the Galactic center, we perform 3D hydrodynamic simulations to trace the complex gaseous flows, and then carry out 3D radiative transfer calculations to compute the variable X-ray spectra. 
  The model WR~140 \textit{RXTE} light curve matches the data well for all phases except the X-ray minimum associated with periastron, 
  while the model spectra agree with the \textit{RXTE} hardness ratio and the shape of the \textit{Suzaku} observations throughout the orbit.
  The Galactic center model of the \textit{Chandra} flux and spectral shape match well in the region $r$\,$\le$\,$3$", but the model flux falls off too rapidly beyond this radius. 
\end{abstract}
\end{center}

\begin{multicols}{2}

\section{Introduction}
The supersonic speeds and high mass-loss rates of WR winds lead to their collisions generating strong signatures of thermal X-ray emission.
These observations, which have been performed by a wide variety of X-ray telescopes, yield important information about these WR winds since they cause both the X-ray emission and absorption.  To disentangle the physical properties that lead to the observed thermal X-ray observations, we use 3D hydrodynamic simulations to determine the complex density and temperature structure of the interacting winds, and then perform 3D X-ray radiative transfer calculations to match the model X-ray emission to the observations, refining the models if necessary.  
Here we present our work on WR~140 and the Galactic center.

\section{WR~140}

WR~140 is the canonical long-period, highly eccentric, colliding wind binary (CWB).  The binary's X-ray flux \citep{CorcoranP11} show strong variation locked to the orbital period; the flux is $\sim1/$(separation) as expected for adiabatic shocks \citep{StevensBlondinPollock92}, except around periastron where the flux drops and the spectra hardens.

\begin{figure}[H]
\begin{center}
\includegraphics[trim={0 8cm 0 0}  ,clip,width=0.3846\columnwidth]{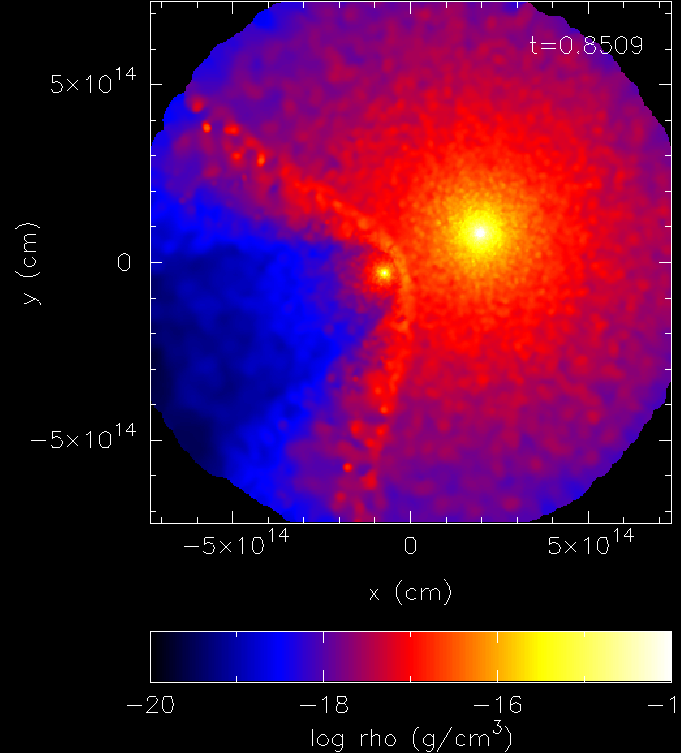}%
\includegraphics[trim={5cm 8cm 0 0},clip,width=0.3048\columnwidth]{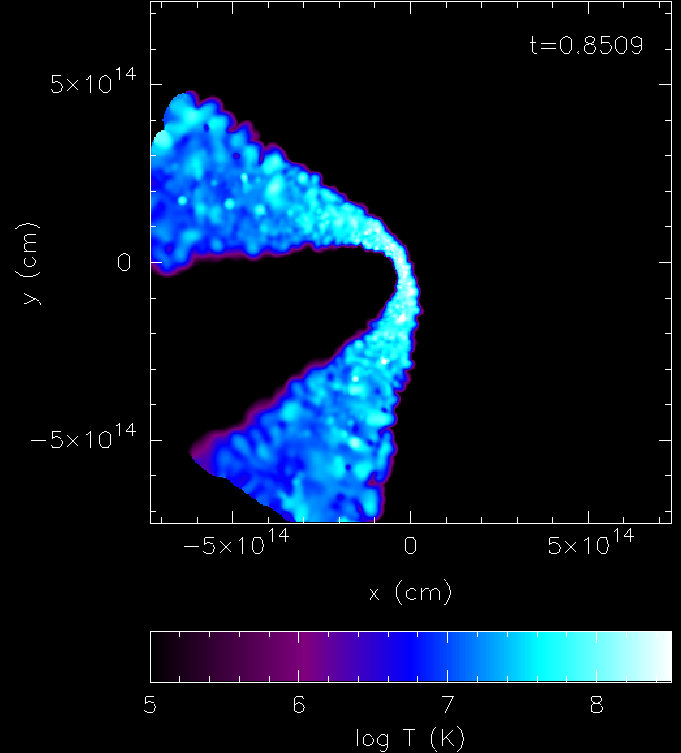}%
\includegraphics[trim={5cm 8cm 0 0},clip,width=0.3048\columnwidth]{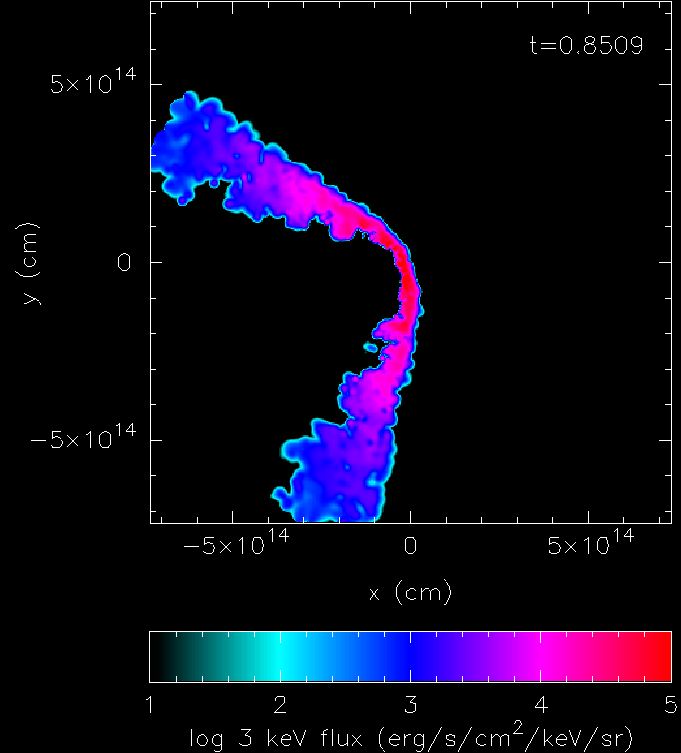}%
\vspace{-0.3mm}
\includegraphics[trim={0 0 0 0}    ,clip,width=0.3846\columnwidth]{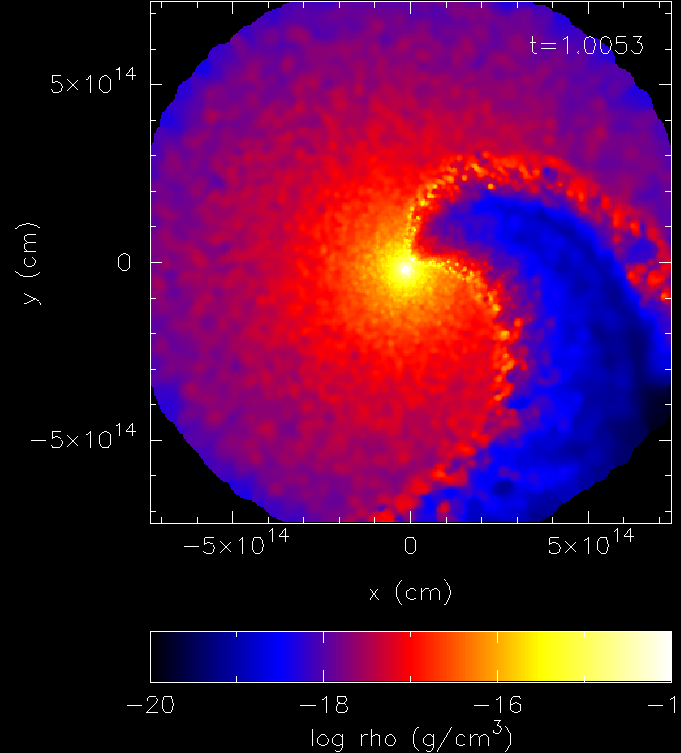}%
\includegraphics[trim={5cm 0 0 0}  ,clip,width=0.3048\columnwidth]{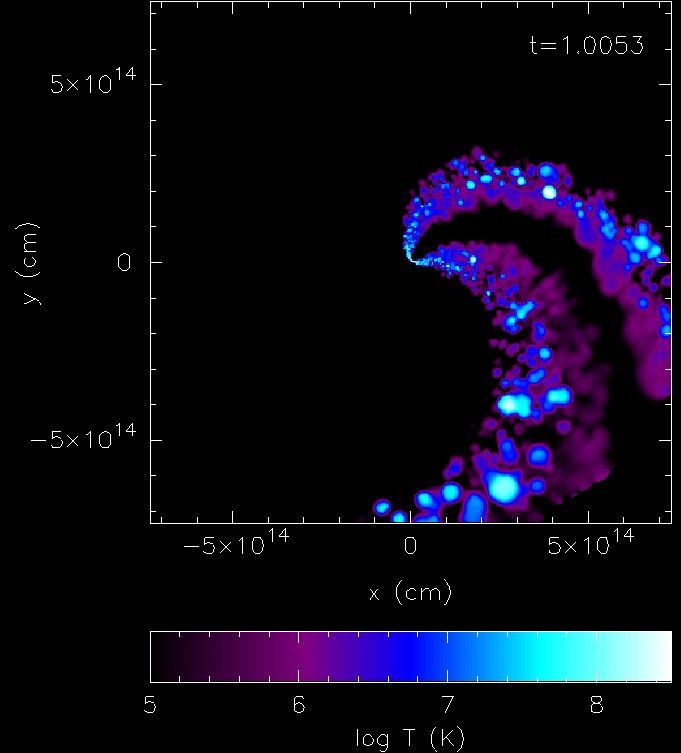}%
\includegraphics[trim={5cm 0 0 0}  ,clip,width=0.3048\columnwidth]{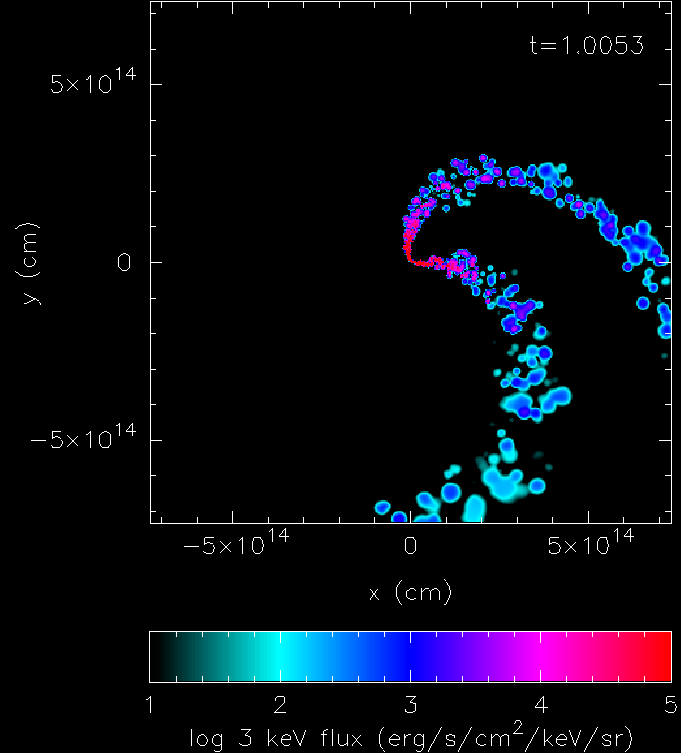}
\caption{WR~140 density (left), temperature (center), and 3 keV X-rays (right) in the orbital plane at phase 0.85 (top) and 0.005 (bottom).
\label{WDens}}
\end{center}
\end{figure}

Fig.\ \ref{WDens} shows the density, temperature, and 3~keV X-ray emission in the orbital plane of the 3D smoothed particle hydrodynamics (SPH) simulation of WR~140.  The winds accelerate from their stellar

\noindent surfaces according to a $\beta$\,=\,1 velocity law, $v(r)=v_\infty(1-R/r)$.  The abundances of the two winds, which enter the SPH simulations via the mean molecular weight and the radiative cooling rates from \texttt{Spex} \citep{SchureP09}, are from \citet{AsplundP09}  for the O4-5 star and from \citet{Crowther07} for the WC7 star, namely $X_{\rm He}=0.6$, $X_{\rm C}=0.31$, and $X_{\rm O}$=0.07.  The radiative transfer calculation also uses the abundances for the X-ray emissivites, which come from the \texttt{APEC} model \citep{SmithP01}\hfill in\hfill \texttt{XSpec}\hfill \citep{Arnaud96},\hfill and\hfill the\hfill wind

\begin{figure}[H]
\begin{center}
\includegraphics[trim={0 0 0 0.2cm},clip,width=\columnwidth]{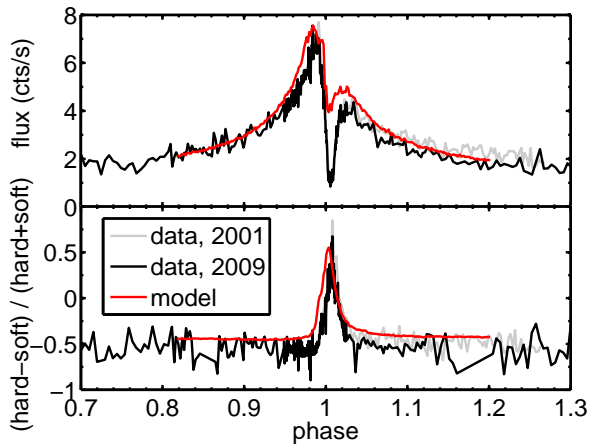}
\caption{\textit{RXTE} light curve (top) and hardness ratio (bottom) comparing the data when periastron occurred in 2001 (gray), in 2009 (black), and the model (red).
\label{WRXTE}}
\end{center}
\end{figure}

\begin{figure}[H]
\begin{center}
\includegraphics[trim={0 0 0 0.3cm},clip,width=\columnwidth]{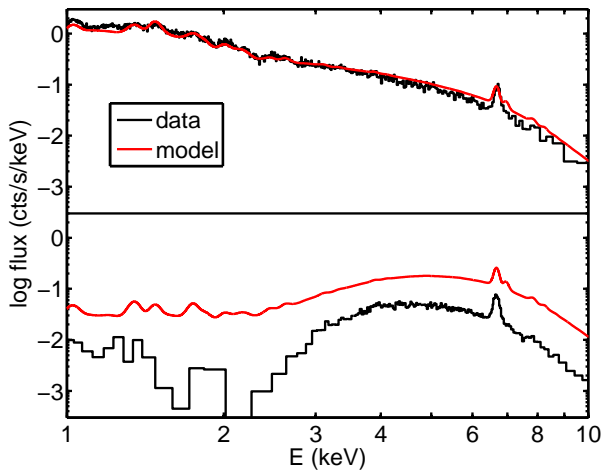}
\caption{\textit{Suzaku} XIS spectra at phase 0.904 (top) and 1.000 (bottom).  The observations are from \citet{SugawaraP15}.
\label{WSpec}}
\end{center}
\end{figure}

\noindent opacities, which come from \texttt{windtabs} \citep{LeuteneggerP10}.  The ISM opacities are from \texttt{TBabs} \citep{WilmsAllenMcCray00},
and the basis of the radiative transfer calculation is the SPH visualization program \texttt{Splash} \citep{Price07}.

Fig.\ \ref{WRXTE} shows the \textit{RXTE} 2-10 keV light curve and hardness ratio comparing the 7.5 keV and 3 keV channels, while Fig.\ \ref{WSpec} shows the \textit{Suzaku} spectra on the rise to maximum and at periastron.
The model X-ray flux, in absolute units, matches well for the majority of the orbit, but does not decrease enough around periastron.
On the other hand, the hardness ratio and spectral shapes match well throughout the orbit, so a gray reduction in flux at periastron is needed to improve the models.

One possibility is to reduce the O-star wind around periastron so it carves out a smaller cavity in the WC wind, thus making less of the WC wind shock.  This will not alter the pre-shock speed of the WC wind, but will reduce its X-ray flux at all energies.  \citet{ParkinSim13}, based on this phenomena occurring in X-ray binaries \citep{StevensKallman90}, explored how wind-wind collision X-rays could reduce the wind strength by ionizing its acceleration region, 
though the colliding-wind X-rays are not strong enough to ionize the O-star wind in WR~140.  Alternatively, the hot radiation from the WC star itself could provide the source of the ionizing photons, which will be explored in future work.


\section{Galactic center}

\citet{CuadraNayakshinMartins08} used the SPH code \texttt{Gadget-2} to follow the orbits of the 30 WR stars within 12" (1"$\sim$0.04 pc) of the SMBH at the Galactic center from 1100 years ago to the present day, all while these stars are ejecting their wind material. 
The simulation volume quickly fills up to form an ambient medium of `old' ejected material, into which the `new' ejected material 
creates bowshocks (spherical bubbles) around the fast- (slow-) moving WR stars.  This 
hot post-shocked gas emits thermal X-rays.

Following the \textit{Chandra} X-ray Visionary Project of the Galactic center \citep{WangP13}, \citet{CuadraNayakshinWang15} improved the hydrodynamic simulations by incorporating several SMBH outflow models potentially associated with the radiatively inefficient accretion flow.  These range from an outflow of $v=10,000$km/s and $\dot{M}_\textrm{out}=\dot{M}_\textrm{accrete}$ over the entire simulation time, to an outburst of $v=10,000$km/s and $\dot{M}_\textrm{out}=10^{-4}M_\odot$/yr lasting from 400 to 100 years ago as suggested by X-ray light echo observations \citep{PontiP10}.

\begin{figure}[H]
  \includegraphics[width=\columnwidth]{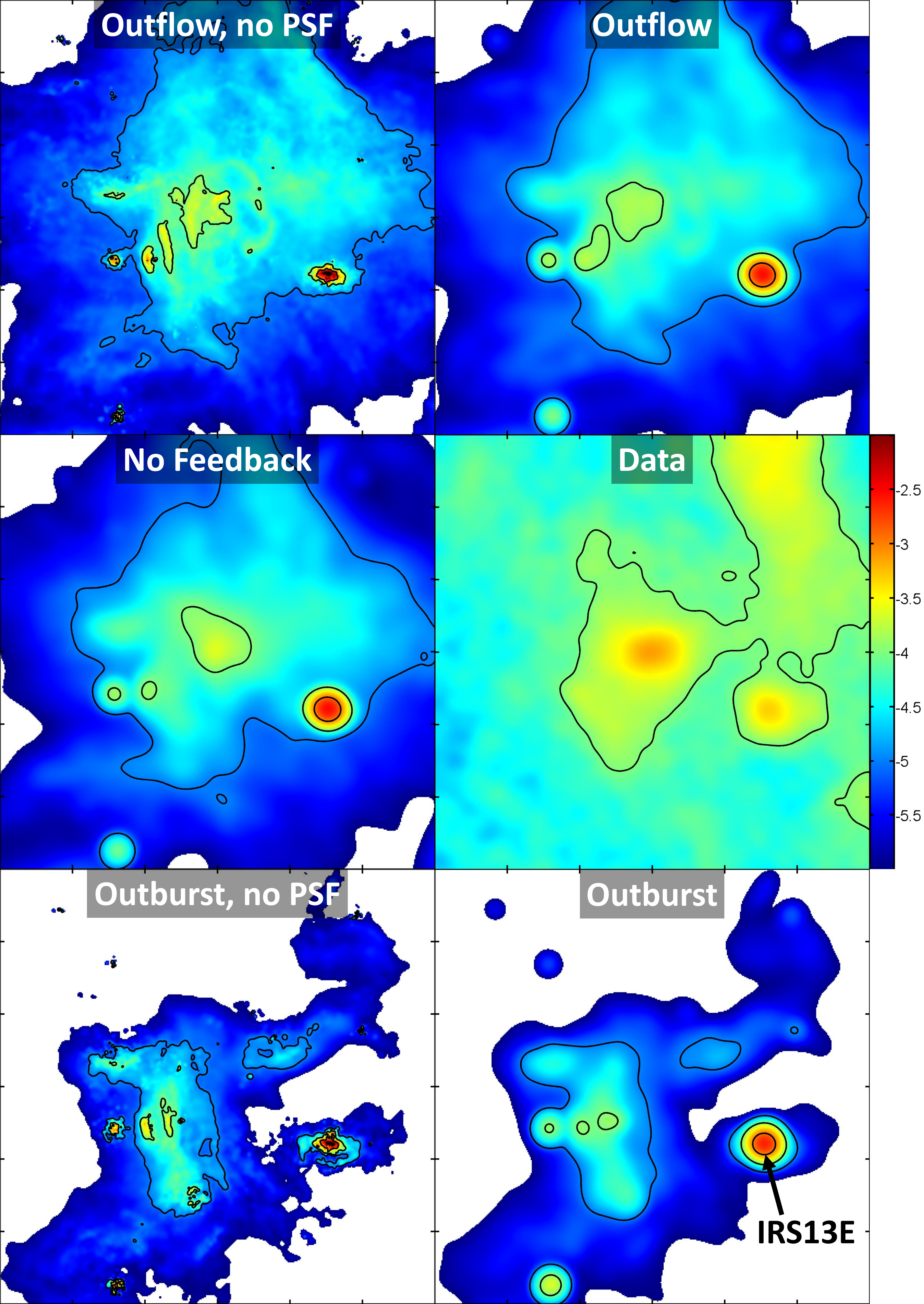}
  \caption{\textit{Chandra} 1-9 keV images ($\pm$6") of the Galactic center with various SMBH feedback models. 
  The images with the PSF folding are directly comparable to the data.  The color bar units are the log of cts/s/arcsec$^2$.  The image color is white below the color bar range.
  \label{GCI}}
\end{figure}

To anchor these simulations in observations, we perform the same X-ray radiative transfer calculation as with WR~140, except that we use the following abundances for the various WR spectral types: the WC8-9 stars have the same WC abundance as the WC7 in WR~140, the WN5-7 stars have WN6 abundances from \citet{Onifer08}, and the WN8-9 stars have WN8 abundances from the \texttt{CMFGEN} website.
Figure \ref{GCI} shows the 1-9 keV \textit{Chandra} ACIS-S HETG 0th order image of the central $\pm$6" for the three models.  To account for the PSF, we fold the model images through a 0.5" FWHM Gaussian.

Since the model does not include the swath from the pulsar wind nebulae in the upper right portion of the plot, nor any point-like emission from the SMBH, the valid region of comparison is the remaining diffuse emission.
The no-feedback  model matches well the region from just beyond the SMBH's influence to $\sim3"$ in radius, but then falls 
off\hfill much\hfill more\hfill quickly\hfill than\hfill the\hfill data\hfill farther\hfill out.\hfill Also,

\noindent there are a few stars whose immediate vicinities have too large an X-ray flux compared to the observations, the brightest of which is the IRS13E cluster, which contains two closely located WR stars.
The outflow model slightly increases the X-ray emission, which is expected since more energy is being added to the simulation, while the outburst model significantly decreases the X-rays through clearing the hot, X-ray-producing gas out of the simulation volume.  It is excluded as a viable feedback model.

\begin{figure}[H]
\begin{center}
  \includegraphics[trim={0 0 0 0.6cm},clip,width=\columnwidth]{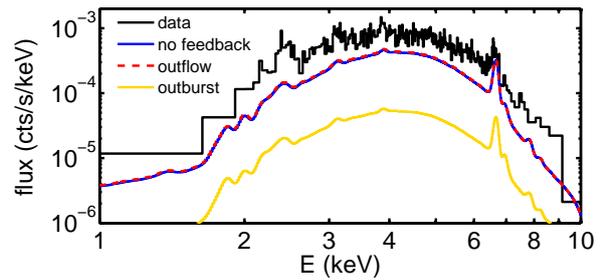}
  \caption{\textit{Chandra} spectra of the Galactic center. 
  \label{GCS}}
\end{center}
\end{figure}

Figure \ref{GCS} shows the X-ray spectra from a 2"-5" ring (excluding IRS13E).  The shape of all feedback models match the data well for $n_H=1.5\times10^{23}$cm$^{-2}$, consistent with the value from analyzing the SMBH emission in the \textit{Chandra} data \citep{WangP13}.

To improve the viable models, the IRS13E flux can be decreased by lowering one or both wind strengths, or increasing the stellar separation.  Increasing the diffuse emission beyond $\sim3$", but not below this radius, is more challenging since increasing mass loss rates will also increase the central X-ray emission, while raising wind speeds will make the spectra harder.  More ambient gas in the outer regions would improve\hfill the\hfill models\hfill since\hfill the\hfill adiabatic\hfill WR\hfill shocks

\noindent will occur closer to their star, increasing their emission according to $\sim1/d_\textrm{shock}$.  Depending on their locations, the SMBH-orbiting O stars might provide this extra gas, and will be included in future work.


\bibliographystyle{aa} 
\bibliography{sampleAF,sampleGM,sampleNS,sampleTZ}

\end{multicols}